\documentstyle[preprint,eqsecnum,aps]{revtex}

\begin{document}
\title{Energy Dependence of Breakup Cross Sections of
Halo Nucleus $^8B$ and Effective Interactions}
\author{C. A. Bertulani$^a$, P. Lotti$^b$, and H. Sagawa$^c$\\
\small $^a$ Instituto de F\'\i sica, Universidade Federal 
do Rio de Janeiro, \\
\small 21945-970 \ Rio de Janeiro, RJ, Brazil. E-mail: 
bertu@if.ufrj.br\\
\small $^b$ Istituto Nazionale di Fisica Nucleare, Sezione di Padova,\\
\small Via Marzolo 8,  35131 Padova, Italy. E-mail: Paolo.Lotti@pd.infn.it\\
\small $^c$  Center for Mathematical Sciences, University of Aizu,\\
\small Aizu Wakamatsu, Fukushima 965, Japan. E-mail: sagawa@u-aizu.ac.jp
}
\maketitle

\begin{abstract}
We study the energy dependence of the cross sections for nucleon
removal
of $^8B$ projectiles. It is shown that the
Glauber model calculations with nucleon-nucleon
t-matrix reproduce well the energy dependence of the
breakup cross sections of $^8B$. 
A DWBA model for the breakup cross section is also proposed and
results are compared with those of the
Glauber model. We show that to obtain an agreement between
the DWBA calculations, the Glauber formalism, and 
the
experimental data,
it is necessary to modify the energy behavior of  
the effective interaction. In particular, the breakup 
potential has a quite
different  energy dependence than the strong absorption potential.
\end{abstract}


\section{Introduction}

Study of breakup reactions with halo nuclei 
is one of the main tools to understand
their structure. The measurements of the width of 
momentum distributions of fragments, the magnitude of the
total reaction cross sections, and single- and double-nucleon removal cross
sections have been of major usefulness to unravel their internal properties
(for a review, see, e.g., \cite{HJJ95}).
These measurements have been carried out at relatively high energies, in the
beam energy range of 30 MeV/nucleon to 1.2 MeV/nucleon.

The Glauber formalism is the major theoretical approach in use to analyse
these measurements. This formalism is well established and yields
very reasonable results for the reactions involving stable 
nuclei at high energies. In particular, a direct connection of the 
quantum mechanical breakup amplitudes and semiclassical calculations
can be done in the Glauber formalism in a very intuitive
way \cite{Gla58}.

In perturbation theory the transition amplitude is given by
\begin{equation}
T_{fi} =\left\langle \phi_f \; \Psi^-\left|U\right|\phi_i\; \Psi^+
\right\rangle \ ,
\label{amp1}
\end{equation}
where $<\phi_i| \; (<\phi_f|)$ denotes the initial (final) 
internal wavefunction
of the nuclei, $\Psi^-\; (\Psi^+)$ is the incoming (outgoing) scattering
wave of the center of mass, and $U$ is the
interaction potential. The Glauber formalism uses eikonal wavefunctions
for the scattering waves. The product $\Psi^{-*}
 .  \Psi^+$ is then simply a
plane wave displaced by a (eikonal) phase which is directly proportional to the
integral of the absorptive potential along the beam direction, the 
$z$-axis. 

The use of eikonal wavefunctions is a crucial step in the Glauber
formalism. Indeed, as shown by Glauber in his excellent lecture
notes \cite{Gla58}, the $z$-integration can be done by parts, and the
potential U in eq. (1) will only appear in the exponential phase. This
procedure is valid if the excitation potential $U$ is the same as the
absorptive potential appearing in the eikonal phases. This is the case
for example in the calculation of total reaction cross sections. However,
it is not a general premise. For example, in inelastic excitations of
surface vibrations, the absorptive potential, $U_{abs}$, causing the 
phase-shifts in the elastic channel is not the same as the interaction 
potential $U_{ex}$ which leads to surface vibrations, although 
(in some models) they  can be related  by derivatives. 
But, even in such situations, the energy dependence
of both potentials are roughly the same. However, while the excitation
potential is related to a few  reaction channels, the
absorptive potential carries information of {\it all} channels which
may lead to the absorption of the scattering waves. Thus, one
expects that a difference in the energy dependence of
the interaction and the absorptive should be manifest in
some sensitive cases. 

A good place to look for a
deviation from the Glauber theory is the breakup reactions involving
halo nuclei. This is because the energies involved in the breakup
are basically the separation energies of the valence nucleons while the
core nucleons which are also relevant for the absorptive part of the
potential have much larger separation energies. Also, the spatial
distribution of the valence and core nucleons are very different so that
they influence
differently on the absorptive (for which all
nucleons participate)  
and the excitation (for which only the valence nucleons participate)
potential.

In section 2 we illustrate the connection of the Glauber formalism and
the nucleon-nucleon cross sections. We apply the theory to the
calculation of proton removal cross sections from $^8B$ projectiles
at several bombarding energies.
In section 3 we develop a DWBA formalism for the stripping reactions.
In section 4 we show how to relate the DWBA calculation with the
Glauber formalism, using effective interactions. Our conclusions are
presented in section 5.  
     
\section{Glauber model and nucleon-nucleon scattering
amplitudes}

In the Glauber theory, after the z-integration, the remaining integrals 
in eq. (1) can be easily related to
the concept of  impact parameter and to absorption and survival 
probabilities.  
For example, simple manipulations show that the 
nucleon removal cross sections in high energy collisions are
described in the Glauber theory by
\begin{equation}
\sigma= 2 \pi \int db \; b\;
\left[ 1-\exp\left(-2{\cal I}m \chi_v
\right)\right] \; \exp\left(-2{\cal I}m \chi_c \right)\; ,
\label{sig1}
\end{equation}
where ${\cal I}m$ stands for the imaginary part,
$v\ (c)$ denotes for valence (core) particles, and $\chi$
are the eikonal phases given by
\begin{equation}
\chi_i(b)= 
-{1 \over \hbar v} \; \int_{-\infty}^\infty dz' \; U(r') \; ,
\label{chi1}
\end{equation}
where $v$ is the projectile velocity,
$r'=\sqrt{b^2+z'^2}$, and $U$ is the optical potential
for the system composed of the particle $i\ (=v,\ c)$ and the target. 
The
term inside the  brackets in eq. (2) can be
interpreted as the probability that the valence nucleons will
be removed in a collision with impact parameter $b$, while the
exponential term outside brackets is the probability that the
core nucleons will survive. This product integrated over all
impact parameters gives the cross section for (valence) nucleon
removal.  

A great simplification introduced by Glauber was to relate the
optical potentials to the nucleon-nucleon cross sections. This can
be done easily by noticing that the optical theorem 
for the
forward nucleon-nucleon amplitude yields
\begin{equation}
t_{NN}({\bf q}=0)=-{4\pi\hbar^2 \over 2 \mu} \; 
f(\theta=0^\circ )=-i 
{\hbar v \over 2} \; \sigma_{NN}(E)\; \left[ 1-
i\alpha (E)\right] \; ,
\label{tnn}
\end{equation}
where $\sigma_{NN}$ is the nucleon-nucleon cross section and 
$\alpha$ is the real-to-imaginary ratio of the nucleon-nucleon
scattering amplitude. 
With the assumptions that    
only very forward angles are involved, and that the nucleon-nucleon
interaction is of very short range (i.e.,
a delta function interaction), 
one can construct optical potentials for the nuclear scattering in
terms of the folding integrals  
\begin{equation}
U_i(R) = t_{NN} ({\bf q}=0) \int \rho_i({\bf r}) \rho_A({\bf R-r}) \;
d^3r \; ,
\label{ui}
\end{equation}
where $\rho_i$ and $\rho_A$ are the ground state densities of the
projectile $i$, and the target $A$, respectively. 

As an application of this model, let us consider the proton removal
cross sections of $^8B$ projectiles in reactions with carbon targets.
For the valence nucleon we get the
density distribution from a Woods-Saxon+spin-orbit potential
model for a proton
in the $p_{3/2}$ orbital (the parameters are given in section 4). 
For the core ($^7Be$) density we use the ground state
density parameterized as $\rho(r)=\rho_0 \left[ 1 +cr^2/a^2\right] \;
\exp\left(r^2/a^2\right)$, with $a=1.77$ fm and $c=0.327$ fm.
The result of the calculation is shown by the short-dashed line of
figure 1. The experimental data are from ref. \cite{Bla97}. 
Although the magnitude
of the cross section is a little overestimated, we see that the
energy dependence follows very closely that of the experimental data.
The dashed curve is the calculation renormalized to the
lowest experimental data point.
In fact, the reasonably good agreement between the energy 
dependence deduced from the  Glauber theory and 
the experimental data on the total
nuclear cross sections, and nucleon removal cross sections, is well
established, both for stable and unstable nuclei. 

Also shown in figure 1 (solid curve) is the model developed
by Hansen \cite{Han96}. 
In his model, the nucleon removal cross section is
forced to have the same energy dependence as the total reaction
cross section. The total reaction cross section has a slightly
different energy dependence than the
valence nucleon removal cross section. This  
can be best seen from the Glauber 
theory. The calculation of
the total reaction cross section
amounts in replacing the integrand in eq. (2) by
$b \; \left[1-\exp\left(-2{\cal I}m \chi_{aA}\right)\right]$
where now $\chi_{aA}$ is the eikonal phase for the collision of 
the projectile $a$ and the target $A$.
In fact, we see that the Hansen's model predicts a rather different
energy dependence of the proton removal cross section. The
data favor the calculation following eq. (2).

From a general point of view, the energy dependence of the
total and the nucleon removal cross sections are directly related
to the underlying optical potentials for the reaction. From the
above discussion we can see that these optical potentials 
should have a similar energy dependence as the 
nucleon-nucleon cross section. 
To study this idea further, let us formulate a DWBA model for
the breakup cross section. 
The use of an effective nucleon-nucleon interaction, M3Y interaction,
will serve as guide 
to understand the link between the optical
potentials and the nucleon-nucleon cross sections.

\section{DWBA breakup amplitudes}

Let us consider the general case of the stripping of the projectile $a$
incident on target $A$:
\begin{equation}
a+A \longrightarrow b \; (a-x) + B\; (A+x) \; .
\label{aa}
\end{equation}
The hamiltonian for the system is
\begin{equation}
H=H_a+H_A+T_{aA}+V_{aA}=H_b+H_B+T_{bB}+V_{bB} \; ,
\label{h}
\end{equation}
The transition matrix element for this reaction is
given by  
\begin{equation}
T=\left\langle\Psi^-({\bf k}_b, {\bf r}_{bB})
\phi_b({\bf s}_b)\phi_B(\xi, {\bf s}_x, {\bf r}_{xA})
\left|U_{bu}\right|
\Psi_a^+({\bf k}_a,{\bf r}_{aA}) 
\phi_a({\bf s}_b, {\bf s}_x, {\bf r}_{bx}) \phi_A(\xi) 
\right\rangle \ ,
\label{t}
\end{equation}
where $\phi_a$, $\phi_b$, $\phi_A$, $\phi_B$ are eigenstates of
$H_a$, $H_b$, $H_A$, $H_B$, respectively, i.e., $H_A\phi_A=
\epsilon_A\phi_A$, $H_a\phi_a=\epsilon_a\phi_a$, etc., $\Psi^\pm$
are distorted waves of the particles $a$ and $b$, i.e., 
$(T_{aA}+U_{aA})\Psi_a^\pm=(E-\epsilon_a-\epsilon_A)\Psi_a^\pm$.
The internal coordinates of $a, \ b$ and $x$,
respectively, are denoted by
${\bf s}_i$ ($i=a,\ b,\ x$), ${\bf r}_{ij}$ are the relative coordinates of
particles $i$ and $j$, and $\xi$ is the internal coordinate of
particle $A$. 

We use the coordinate relationship
\begin{equation}
{\bf r}_{bB} =  {\bf r}_{bx}+\left( {m_A\over m_B} \right) \; 
{\bf r}_{xA} \ , \ \ \ \ \  
{\bf r}_{aA} =  {\bf r}_{xA}+\left( {m_b\over m_a} \right) \; 
{\bf r}_{bx} \ ,
\label{rbb}
\end{equation}
and we integrate over the internal coordinate, $\xi$,  of $A$
\begin{equation}
\psi_x({\bf s}_x, {\bf r}_{xA}) = \int d\xi\ \phi_B^*(\xi,
\ {\bf s}_x, {\bf r}_{xA}) \; \phi_A(\xi) \ ,
\label{psix}
\end{equation}
and over the internal coordinates of $b$ and $x$,
\begin{equation}
\int d{\bf s}_b \ d{\bf s}_x \ \phi_b^*({\bf s}_b)
\; \psi_x \left({\bf s}_x, {\bf r}_{xA}\right) \
\phi_a({\bf s}_b, {\bf s}_x, {\bf r}_{bx})= 
C_{bx}\;  \phi_a(
{\bf r}_{bx})\; \Psi_x({\bf r}_{xA}) \ ,
\label{intds}
\end{equation}
where $|C_{bx}|^2$ is the spectroscopic factor.

We get for the transition matrix element
\begin{eqnarray}
T&=&C_{bx}\int d^3 r_{bx} \; d^3 r_{bA}
\ \Psi_b^{-*}\left({\bf k}_b, {\bf r}_{bx}+
{m_A\over m_B}{\bf r}_{xA}\right)
\phi_x({\bf r}_{xA}) 
\nonumber \\
&\times&
U_{bu} ({\bf r}_{xA}, \; {\bf r}_{bx})\ \phi_a({\bf r}_{bx})
\Psi_a^+\left({\bf k}_a, {\bf r}_{xA}+
{m_b\over m_a}{\bf r}_{bx}\right) \ .
\label{tcbx} 
\end{eqnarray}
 
The potential $U_{ex}$ for the breakup channel is given by
\begin{eqnarray}
U_{bu} ({\bf r}_{xA}, \; {\bf r}_{bx})\
&=& U_{bA}({\bf r}_{bA})+U_{xA}({\bf r}_{xA})-
U_{aA}({\bf r}_{aA})  \nonumber \\
&=&U_{bA}({\bf r}_{xA}+{\bf r}_{bx})+U_{xA}({\bf r}_{xA})-
U_{aA}\left({\bf r}_{xA}+{m_b\over m_a}{\bf r}_{bx}\right) \ . 
\end{eqnarray}
 
If we now integrate over ${\bf r}_{bx}$, we can use the fact that
the bound state wavefunction is peaked at small ${\bf r}_{bx}$
values, so that
\begin{eqnarray}
&&\int d^3 r_{bx}\; 
\Psi_b^{-*}\left({\bf k}_b,{\bf r}_{bx}+
{m_A\over m_B}{\bf r}_{xA}\right)
U_{bu}({\bf r}_{xA}, {\bf r}_{bx}) \phi_a({\bf r}_{bx})
\Psi_a^+\left({\bf k}_a, {\bf r}_{xA}+
{m_b\over m_a}{\bf r}_{bx}\right)  
\nonumber \\
&\approx& 
\Psi_b^{-*}\left({\bf k}_b, 
{m_A\over m_B} {\bf r}_{xA}\right) 
\Psi_a^+\left({\bf k}_a, {\bf r}_{xA}\right) 
\;
\int d^3 r_{bx}\; 
U_{bu}({\bf r}_{xA}, {\bf r}_{bx}) \phi_a({\bf r}_{bx})
\ 
\label{intrbx} 
\end{eqnarray}

We now define a ``transition", or ``excitation",  potential as
\begin{equation}
U_{ex}({\bf r}_{xA})= \int
d^3 r_{bx} \; \phi_a({\bf r}_{bx})
\; U_{bu}({\bf r}_{xA}, \ {\bf r}_{bx})
 \ ,
\label{ubu} 
\end{equation}
so that
\begin{equation}
T = C_{bx} \; \int d^3r_{xA}\; \Psi_b^{-*}\left({\bf k}_b, \; 
{m_A\over m_B} {\bf r}_{xA}\right) \
\Psi_x({\bf r}_{xA})\; U_{ex}({\bf r}_{xA})\;
\Psi_a^+\left({\bf k}_a, \ {\bf r}_{xA}\right) 
 \ .
\label{tf} 
\end{equation}
The above equation is our main result. It gives the t-matrix in 
terms of the scattering waves of particle $a$, $b$, and $x$,
and a ``transition" potential $U_{ex}$. This potential contains the
information on the structure of particle $a$.

If we are only interested in particle $b$, assuming that the
particle $x$ is not observed, we can use the closure relation
\begin{equation}
\sum_{{\bf k}_x}\Psi_x^{({\bf k}_x )*}({\bf r}_{xA})\;
\Psi_x^{({\bf k}_x )}({\bf r'}_{xA})=\delta\left({\bf r}_{xA}-
{\bf r'}_{xA}\right) 
 \ ,
\label{closure} 
\end{equation}
to obtain
\begin{equation}
\sum_{{\bf k}_x}\left|T\right|^2=
\left|C_{bx}\right|^2 
\; \int d^3r_{xA}\; \left|\Psi_b^{-*}\left({\bf k}_b, \ 
{m_A\over m_B} {\bf r}_{xA}\right)\right|^2 \
\left| U_{ex}({\bf r}_{xA})\right|^2\;
\left|\Psi_a^+\left({\bf k}_a, \ {\bf r}_{xA}\right)
\right|^2 
 \ , 
\end{equation}
The calculation becomes very transparent if we use eikonal
functions for the distorted waves:
\begin{eqnarray}
\Psi_b^-&=&\exp\left[i{m_A\over m_B}{\bf k}_b . {\bf r}_{xA}
+i\chi_b\left({m_A\over m_B} b_{xA}\right)\right]
\nonumber \\
\Psi_a^+&=&\exp\left[i{\bf k}_a . {\bf r}_{xA}
+i\chi_a\left( b_{xA}\right)\right]
 \ ,
\label{eik} 
\end{eqnarray}
with the eikonal phases given by
\begin{eqnarray}
\chi_b\left({m_A\over m_B} b_{xA}\right)
&=&
-{1\over {\hbar v_b}} \int_{z_{xA}}^\infty U_{bA}\left(
{m_A\over m_B} r'_{xA}  \right) dz'_{xA}
\nonumber \\
\chi_a\left( b_{xA}\right)
&=&
-{1\over {\hbar v_a}} \int^{z_{xA}}_{-\infty} 
 U_{aA}
\left(
r'_{xA}  \right) dz'_{xA} 
 \ ,
\label{eik2} 
\end{eqnarray}
 where $r'_{xA}=\sqrt{b_{xA}^2+{z'}_{xA}^2}$. Note that, since only the moduli 
of the $\Psi$'s enter into eq. (18), we get
\begin{equation}
\sum_{{\bf k}_x}\left|T\right|^2=
\left|C_{bx}\right|^2 
\; \int d^3r_{xA}\; S_a(r_{xA})\; S_b(r_{xA})
\; \left|U_{ex}(r_{xA})\right|^2\ , 
\label{sumt} 
\end{equation}
where 
\begin{eqnarray}
S_b(r_{xA})&=&\exp\left[{2\over \hbar v_b}
\; \int_{z_{xA}}^\infty {\cal I}m \; U_{bA}\left(
{m_A\over m_B} r'_{xA} \right) dz'_{xA}\right]
\nonumber \\
S_a(r_{xA})&=&\exp\left[{2\over \hbar v_a}
\; \int^{z_{xA}}_{-\infty} {\cal I}m U_{aA}\left(
r'_{xA} \right) dz'_{xA}\right]
 \ , 
\end{eqnarray}

Finally, the  cross section for removal of particle $x$ from $a$
is given by 
\begin{equation}
\sigma={m_a m_b \over \pi \hbar^4} \; {k_b\over k_a}\;
{\sum_{{\bf k}_x,\ spins} |T|^2 \over (2J_A+1)(2J_a+1)}
\ ,
\label{sigma} 
\end{equation}
and we shall assume that $v_b \approx v_a$, valid for high energy
collisions and small binding energies of the incident projectile.

\section{Wavefunctions and effective interactions}

The ground state wavefunction of  $^8B$, in a given magnetic
substate, $M$, is taken as
\begin{equation}
\phi_a^{(M)}({\bf r}_{bx})= \sum_{m,M_A} \left\langle
jmI_AM_A|JM\right\rangle \; \phi_{jm}({\bf r})\; 
|I_AM_A\rangle
\ ,
\label{phia} 
\end{equation}
where $|I_AM_A\rangle$ 
is the wavefunction of the $^7Be$ ($I_A^\pi
=3/2^-$), and $\phi_{j,m}$ is the single-particle wavefunction 
of the proton $j^\pi=3/2^-$, coupled to a total angular momentum
$J^\pi=2^+$. Thus, the potential $U_{ex}$ in eqs. (15-17,21) 
 depends
on the initial orientation of $^8B$ and the target,
 what means that eq. (23) carries
an average over the magnetic substates of these nuclei.

Using the
properties of the Clebsch-Gordan coefficients, and the orthogonality
of the core wavefunctions, we get for the spin averaged potential
\begin{equation}
{\cal U}_{ex}\equiv
{1\over (2J_A+1)(2J_a+1)}\; \sum_{spins}
|U_{ex}(r_{xA})|^2=
{1\over 20}\; \int d^3 r_{bx}
{R_j(r_{bx})\over r_{bx}}\; U_{bu}
({\bf r}_{xA}, \ {\bf r}_{bx})
\ , 
\end{equation}
where $R_j(r)/r$ is the radial part of the
single-particle wavefunction $\phi_{jm}$. 
The cross section is
\begin{equation}
\sigma={2 m_b^2\over \pi \hbar^4}\; \left|C_{bx}\right|^2
\; \int
db \; b\; S_{ab}(b)\; {\cal F}_{ex}(b)
\ ,
\label{sigf} 
\end{equation}
where $S_{ab}\equiv S_aS_b$ (we neglect the
small dependence of $S_{ab}$ on $z$) and 
\begin{equation}
{\cal F}_{ex}(b)=\int_{-\infty}^\infty dz \ 
{\cal U}_{ex}\left(\sqrt{b^2+z^2}\right)
\ .
\label{calf} 
\end{equation}

Now we need to determine the
 optical potentials to proceed with the calculation.
Usually these optical potentials are obtained from elastic scattering
experiments. But, for unstable nuclei the situation is
quite different. One generally has to construct 
these optical potentials theoretically
from effective nucleon-nucleon interactions. 
Among these, one of the most popular
is the M3Y interaction, which has been shown 
to work quite reasonably
for elastic and inelastic scattering of heavy ions
at low and intermediate energy collisions \cite{Ber77,Kob84}.

In its simplest form the M3Y interaction is given by two direct
terms with different ranges, and an exchange term represented
by a delta interaction:
\begin{equation}
t(s)=A{e^{-\beta_1 s}\over \beta_1 s}+
B{e^{-\beta_2 s}\over \beta_2 s}+C\delta({\bf s})
\ , 
\label{M3Y} 
\end{equation}
where
A=7999 MeV, $B=-2134$ MeV, $C=-276$
MeV $fm^{3}$, $\beta_1=4$ $fm^{-1}$,
and $\beta_2=2.5$ $fm^{-1}$.
The real part of the optical potential is obtained from a folding of this interaction with the
ground state densities of the nuclei:
\begin{equation}
U_{ij}({\bf R})=\int d^3r_1\; d^3r_2 \; \rho_A({\bf r}_1)
\rho_j({\bf r}_2) \; t(s)
\ ,  
\end{equation}
with ${\bf s}={\bf R}+{\bf r}_2-{\bf r}_1$. 
The imaginary part of the optical 
potential is usually parameterized to be
${\cal I}m U=\lambda U_{M3Y}$, with 
$\lambda = 0.6-0.8$. 

The $M3Y$ interaction (28) has been 
modified to account for the 
energy dependence on the beam energy. However, for the
energy range of $5-50$ MeV/nucleon, only a small
energy dependence was introduced  \cite{Kob84} as 
a variation of  the 
exchange term.

To study the breakup of $^8B$ projectiles,
we will use the form given by eq. (28) 
for the M3Y interaction
with the $^7Be$ and $^8B$ densities as in section 1, and a proton
gaussian density of radius equal to $0.7$ fm. The radial 
wavefunction of the proton, $R_{3/2}$, was obtained in
a Woods-Saxon+spin-orbit potential, i.e., $V(r)=
V_0\left[1-F_{so}({\bf l.s})(r_0/r) d/dr\right]f(r)$, 
with $f(r)=\left[1+\exp\left((r-R)/a\right)\right]^{-1}$
with parameters $V_0=-44.66$ MeV,
$a=0.56$ fm, $r_0=1.25$ fm, $R=2.391$ fm,
$F_{so}=0.351$ fm, which reproduces the binding energy,
$\epsilon=0.138$ MeV, of $^8B$. 
The spectroscopic factor is 
taken as unity.

In figure 2 we plot the function ${\cal F}_{ex}(b)$, which 
contains the information not only of the ground-state wavefunction of the
$^8B$, but also on the effective interaction. In figure 3 we plot the
profile function $S_{ab}(b)$, which depends only on the effective
interaction. We have calculated it for the energies $E/A=$30,
150, 300, 800, and 1200 MeV, respectively. 
The magnitude of the cross section is proportional
the area below $S_{ab}(b)\times {\cal F}_{ex}(b)$. Since 
${\cal F}_{ex}$ does not depend on the beam energy, the energy dependence
is solely due to $S_{ab}$. Since the
M3Y interaction does not depend on the
energy, the energy dependence is a consequence of the
$\hbar v$ factors in the denominators appearing
in eqs. (22). This causes the nuclear 
transparency, described by the
factor $S_{ab}$, to increase for small 
$b$ as the beam energy increases. As a consequence, the 
cross sections {\it increase} with energy. A comparison with the
experimental data in figure 4 (dotted curve) shows the departure of the
calculated cross sections from the experimental data at large
energies.

It is clear that we have to modify the effective interaction in
eq. (29) so as to incorporate the energy dependence. A simple
way to do that is to make (29) have the same energy dependence as
in eq. (4,5).  It should be noticed that the potential
$U_{ex}$ in eq. (16)
is not the same as the potentials appearing in the phase of
the scattering waves, in the sense that it does have neither
the same magnitude, nor the same spatial dependence.
If we take, although it is not necessary,
the same energy dependence as in (4)
we can obtain the effective interaction as
\begin{equation}
t(E,s)=-i{\hbar v\over 2 t_0} \; \sigma_{NN}(E) \;
\left[1-i\alpha(E)\right] \; t(s)
\ , 
\label{tes} 
\end{equation}
where $t_0=421$ MeV is the volume integral of the M3Y interaction.
Note that eq. (30) gives the same removal cross section 
as the M3Y interaction for E=30 MeV.
Inserting this result (eq. 30) in eq. (29), we 
can determine the imaginary part of the
optical potential automatically.

We repeat the calculation for the proton removal cross sections of $^8B$
using the effective interaction (30) in eq. (29) and 
the calculated cross sections by eq. (26) are shown in figure
4 (dashed line). We see that the
energy dependence of the cross section changes drastically and follows more 
closely the trend of the experimental data. 

As mentioned above we 
do not need to assume that the absorption potential and the excitation
potential have the same energy dependence. It
is reasonable to assume that the absorption potential follows the 
receipt of eq. (29), with $t$ given by (30), 
since this has the same energy dependence occurring
in all calculations based on the Glauber formalism for total reaction
cross sections, which are known to agree reasonably 
with the experimental 
data. Thus, we change the excitation potential to adjust its 
energy dependence to the data points. We find out that a simple 
energy dependence of the form 
$t(E,s) \propto E^{-0.25}\; t(s)$, for $E<200$ MeV, and
${\cal U}_{ex}\propto constant\; . t(s)$, for $E\ge 200$ MeV,
reproduces the trend of the experimental data, as we show in
figure 4 by the solid line, with a normalization factor which best
fits the data. 

\subsection{Conclusions}

In summary, it is found that the energy dependence of the 
experimental removal cross sections of $^8B$
can be obtained by  Glauber model calculations with the nucleon-nucleon
t-matrix. We have studied also the relation between the Glauber model and 
the DWBA formalism. 
The DWBA approach to 
nucleon removal cross sections in general agrees with
Glauber calculations if the optical potential in the elastic
channel has the same energy dependence as the breakup potential.
For halo nuclei this is not necessarily
true, as we have shown for
the breakup of the $^8B$ nucleus. This finding
might have important consequences, 
not only for the breakup of halo nuclei but also for their
excitations to bound states. More studies with halo nuclei
are needed in order to clarify the role of effective interactions
in the construction of optical potentials, and of their
connection to nucleon-nucleon scattering amplitudes.

\bigskip\bigskip
\noindent{\bf Acknowledgments}

\medskip

This work was supported in part by
the INFN/Italy, the University of Padova/Italy,
and in part by the
MCT/FINEP/CNPq(PRONEX)/Brazil under contract 
No. 41.96.0886.00. 

\bigskip\bigskip



\bigskip 
{\bf Figure Caption}\\

{\bf Fig. 1} - 
Proton removal cross sections of $^8B$ projectiles on carbon targets
as a function of the incident beam energy.
Data points are from ref. \cite{Bla97}. Solid line is a calculation
based on a model by Hansen \cite{Han96}. Short-dashed line 
is a calculation based on the Glauber model, eq. (2). 
Dashed line is simply the short-dashed line downshifted by a
factor of 0.83.   

{\bf Fig. 2} - 
Excitation function ${\cal F}_{ex}$ (see definition in text; eq. (27)),
as a function of the impact parameter.

{\bf Fig. 3} - 
Nuclear transparency function for the proton removal of $^8B$
projectiles at several energies incident on carbon targets.

{\bf Fig. 4} -
Proton removal
cross sections of $^8B$ projectiles incident on carbon targets 
as a function of the incident energy. 
Dotted line is the result of a DWBA calculation
with the folding potentials with M3Y interaction. For the dashed line 
the effective interaction was taken with the
same energy dependence as the nucleon-nucleon scattering amplitude.
Solid line is the result obtained with a fit for the
energy dependence of the breakup potential, different from
that of the absorption potential. For details, see the text.

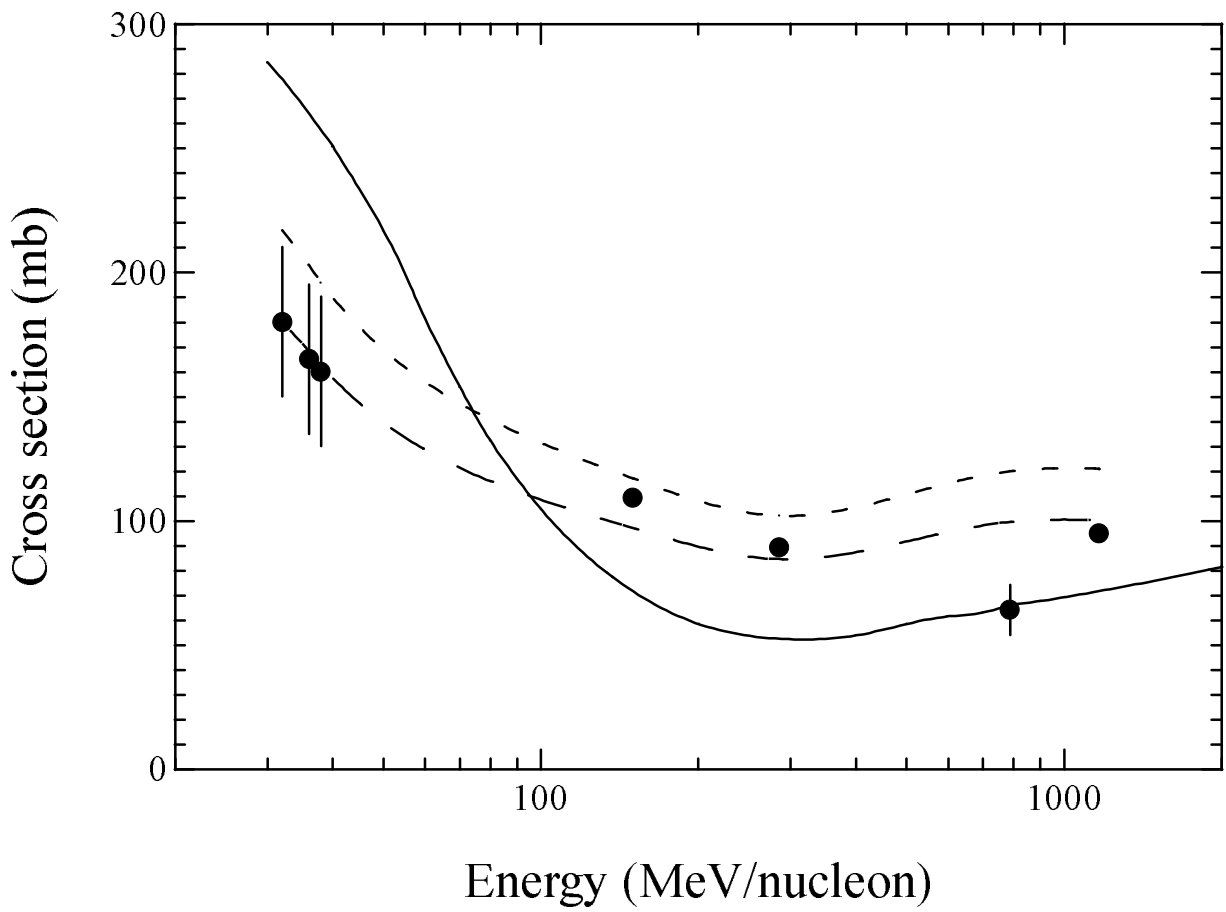
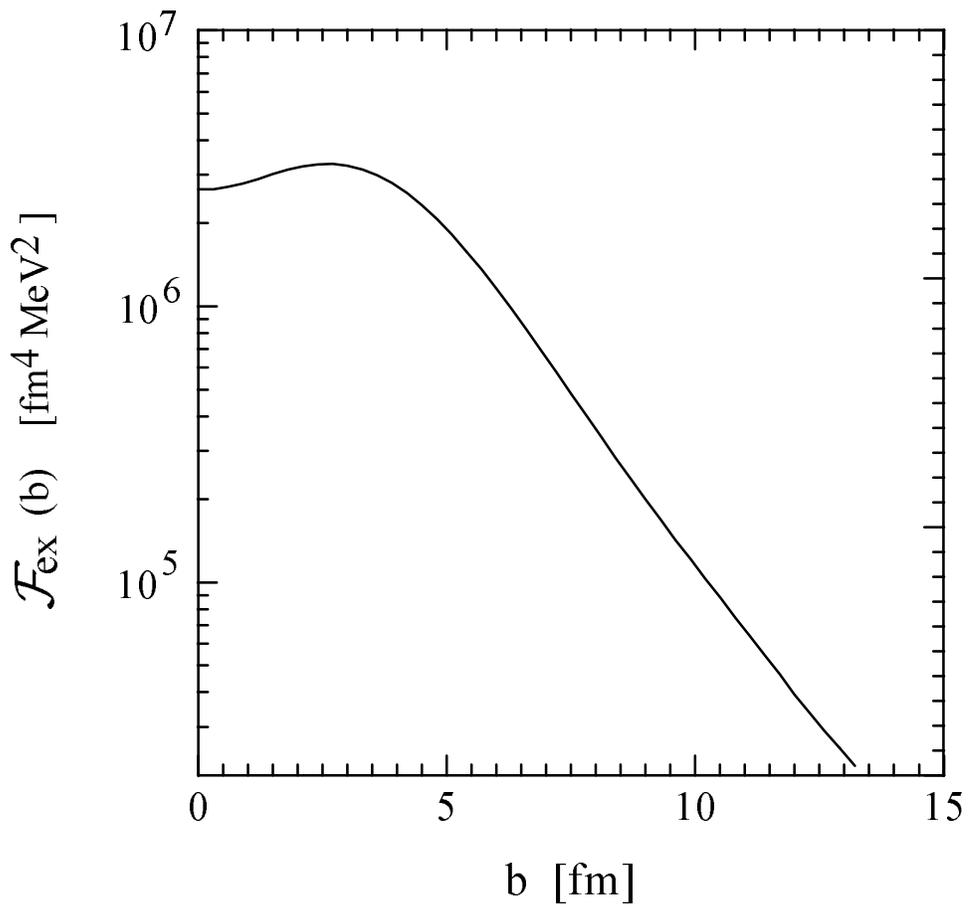
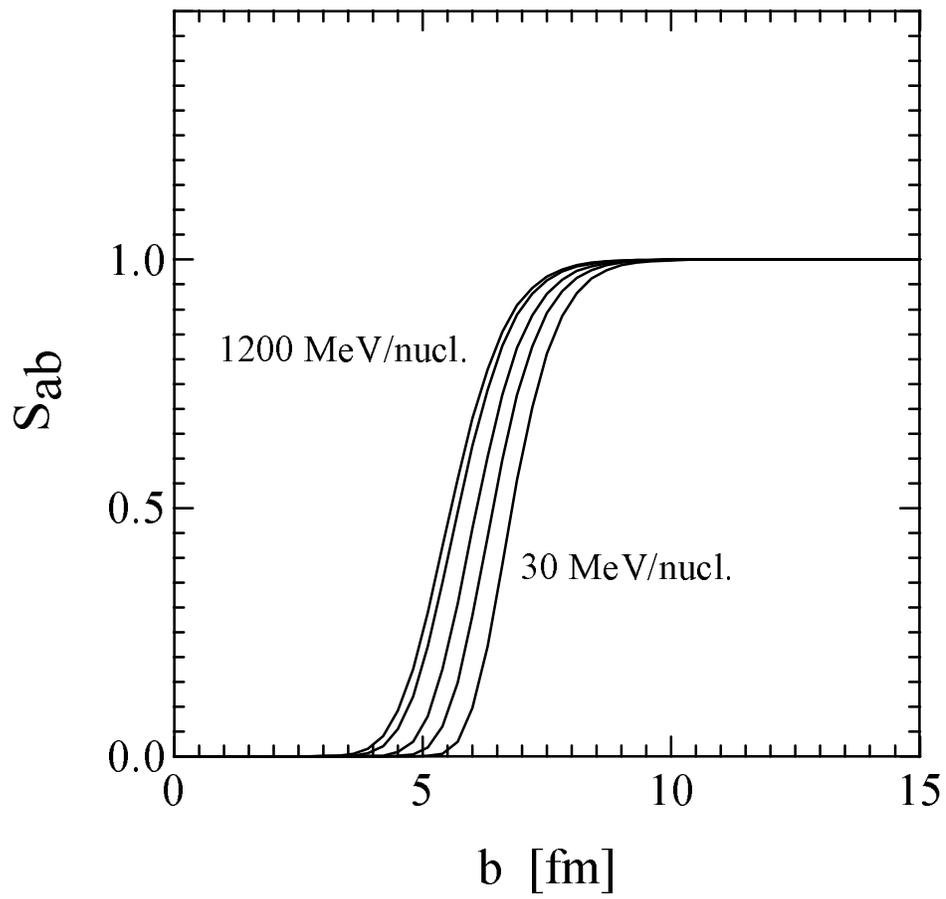
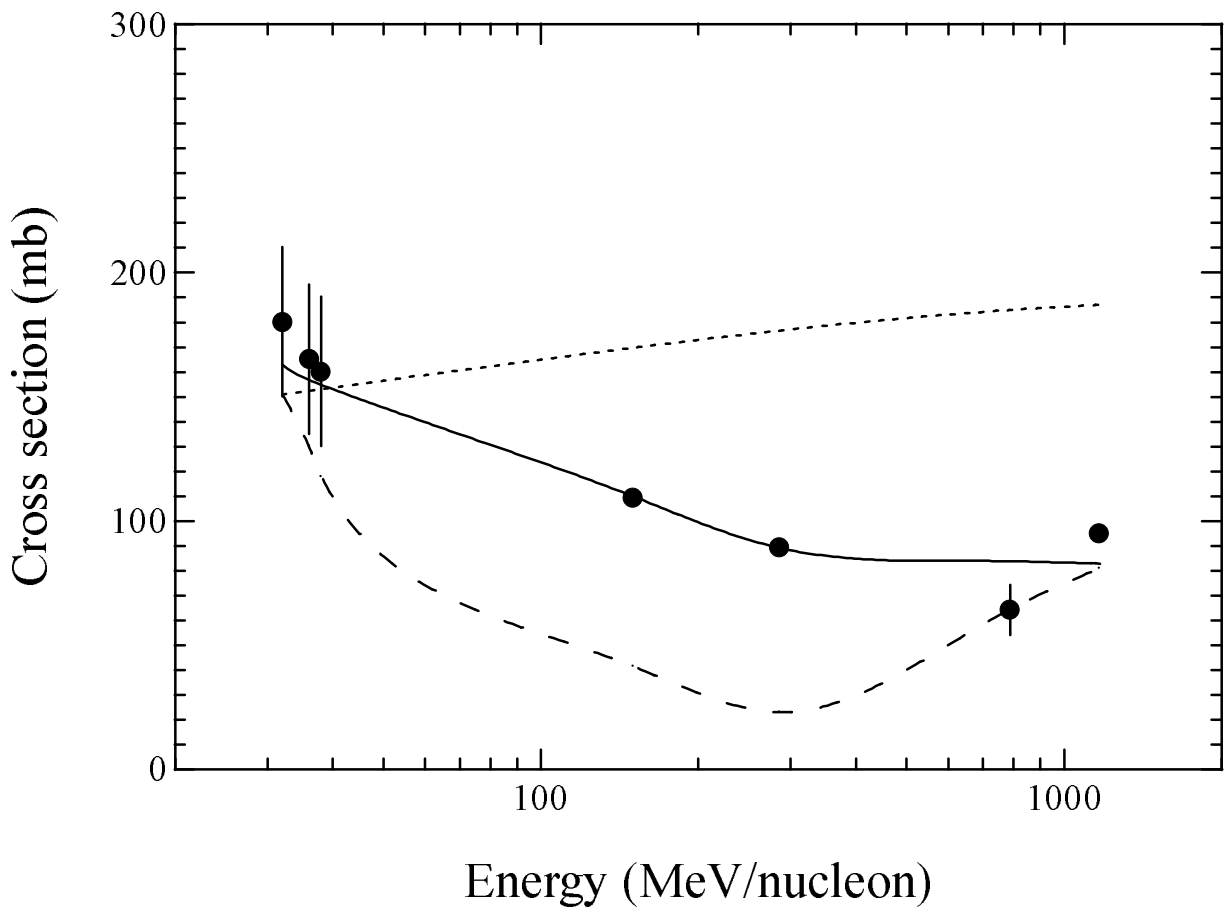
\end{document}